\documentclass[aps,prl,reprint,superscriptaddress,showpacs,showkeys]{revtex4-1}

\usepackage{amssymb}
\usepackage{lipsum}
\usepackage{graphicx}   
\usepackage{dcolumn}   
\usepackage{bm}          
\usepackage{verbatim}  
\usepackage{amsmath}
\usepackage[T1]{fontenc}
\usepackage{mathptmx}
\usepackage{float}

\begin{document}

\title{Emergent kink stability of a magnetized plasma jet injected \\ into a transverse background magnetic field}

\author{Yue Zhang}
\thanks{Present address: University of Washington, Seattle, WA 98195}
\affiliation{University of New Mexico, Albuquerque, New Mexico, 87131 USA}

\author{Mark Gilmore}
\email[]{mgilmore@unm.edu}
\affiliation{University of New Mexico, Albuquerque, New Mexico, 87131 USA}

\author{Scott C. Hsu}
\affiliation{Los Alamos National Laboratory, Los Alamos, New Mexico, 87545  USA}

\author{Dustin M. Fisher}
\affiliation{University of New Mexico, Albuquerque, New Mexico, 87131 USA}

\author{Alan G. Lynn}
\thanks{Present address: Naval Research Laboratory, Washington, DC, 20375}
\affiliation{University of New Mexico, Albuquerque, New Mexico, 87131 USA}

\date{\today}


\begin{abstract}
We report experimental results on the injection of a magnetized plasma jet into a transverse background magnetic field in the HelCat linear plasma device at the University of New Mexico [M. Gilmore et al., J. Plasma Phys. \textbf{81}, 345810104 (2015)]. After the plasma jet leaves the plasma-gun muzzle, a tension force arising from an increasing curvature of the background magnetic field induces in the jet a sheared axial-flow gradient above the theoretical kink-stabilization threshold. We observe that this emergent sheared axial flow stabilizes the $n = 1$ kink mode in the jet, whereas a kink instability is observed in the jet when there is no background magnetic field present. 
\end{abstract}

\maketitle

The collimation and stability of a plasma jet injected into a transverse magnetic field are relevant to astrophysical jets,\cite{honda2002self}~bipolar outflows associated with young stellar~object,\cite{pelletier1992hydromagnetic}~solar-wind evolution,\cite{schatten1969model,jokipii1990polar}~plasma beam focusing,\cite{rosenzweig1991acceleration}~\mbox{magnetotail} physics,\cite{zmuda1970characteristics,lyons2013quantitative}~etc. In magnetic fusion, cross-magnetic-field injection to deliver fuel into the core of a tokamak is necessary for achieving more efficient utilization of \mbox{deuterium--tritium} fuel and optimize the energy confinement time for high fusion gain.\cite{brown1990spheromak,brown1990current,raman1997experimental,olynyk2008development}~The study of plasma motion across a transverse magnetic   field has a long history.\cite{schmidt1960plasma,baker1965experimental}~Modern experimental investigations of plasma jets include those using pulsed-power-driven plasma guns,\cite{hsu2002laboratory,hsu2003experimental,hsu2005jets,bellan2005simulating,yun2007large}~radial wire array~Z~pinches,\cite{lebedev2002laboratory,lebedev2004jet,lebedev2005magnetic,ciardi2007evolution}~and laser-produced plasma.\cite{mostovych1989laser,ripin1990laboratory,harilal2004confinement}~However, nearly all of the experimental studies have been of plasma jets launched into vacuum. In addition, nonlinear ideal magnetohydrodynamic (MHD) simulations have modeled a plasma injected into a transverse magnetic field under varying conditions.\cite{liu2008ideal,liu2009ideal,liu2011ideal}

We report here the experimental observation of emergent kink stabilization of a current-driven plasma jet injected into a transverse background magnetic field.~Experiments have also been carried out in which a plasma-jet is injected into a background plasma~(temperature $T_{e}\sim$ 1-5~eV, density $\sim 10^{17}$~m$^{-3}$), \cite{desjardins2016dynamics,kelly2016ari}~but in such low-$\beta$~($\sim 0.01$) plasmas, the magnetic field provides the dominant mechanism for kink stabilization.~Both charge-coupled device (CCD) camera images and magnetic-field probe data show similar plasma-jet behavior for both magnetic-field-only and plasma cases.~The data presented in this letter are taken from experiments with background magnetic field only.

The plasma jet is formed and launched by a compact magnetized coaxial gun,\cite{zhang2009design}~and the background magnetic field is produced in the Helicon-Cathode (HelCat) linear plasma device at the University of New Mexico.\cite{lynn2009helcat,gilmore2015helcat}~The injection results in several observed phenomena:~$\left(1\right)$~when launching the plasma jet into vacuum, a classical $n = 1$ kink instability is observed, consistent with prior work;\cite{hsu2003experimental,hsu2005jets}~$\left(2\right)$~when injecting the jet into a transverse background magnetic field, a more-stable jet is observed;~and~$\left(3\right)$~analysis of magnetic-field measurements and high-speed-camera images during the injection indicate that the transverse background magnetic field exerts a magnetic tension force on the jet, allowing  the formation of extended-length, kink-free jets. Although, the measured safety factor is below unity, implying kink instability, we show that the stability is a result of an emergent sheared axial flow due to the tension force of the background field acting on the jet.

\begin{figure}
\includegraphics[scale=0.38]{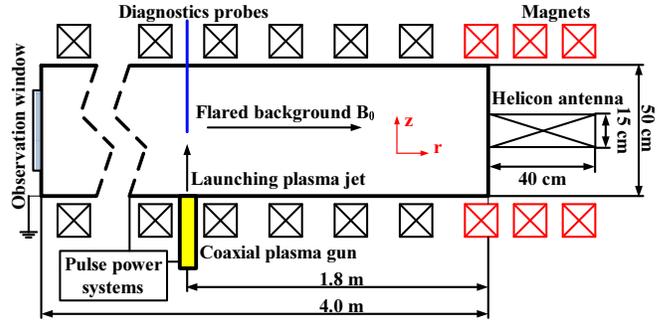}
\caption{(Color online) Top-view schematic of vacuum chamber and coaxial gun setup, not to scale.~A helicon antenna provides the background helicon plasma. Magnetic and electrostatic probes measure localized plasma parameters,~and an intensified CCD camera detects plasma-jet evolution via an end-view window.}
\label{fig:overview}
\end{figure}

HelCat is a 4-m-long and 50-cm-diameter cylindrical device with axial magnetic field, providing a background magnetic field.~The plasma jet is formed by a compact, cylindrical magnetized coaxial plasma gun. As indicated in Fig.~\ref{fig:overview}, the gun is mounted on a side port 1.8~m away from   the helicon source, with an angle of $90^{\circ}$ with respect to the background magnetic field. The center electrode disk is 2.54~cm in diameter, and the cylindrical outer electrode is 4.75~cm in diameter. A fast gas valve puffs neutral argon gas into the annual electrode gap with a peak pressure of 0.25~Torr. An ignitron-switched 10-kV, 120-$\mu$F ($\sim$10-$\mu$s rise time) capacitor bank is employed to fully ionize the neutral gas via the discharge current. An initial radial current sheet is formed between the inner and outer electrodes. The resulting axial $\mathit{J\times B}$ force accelerates the current sheet along the gun axis, forming a plasma jet that propagates into the main vacuum chamber.

\begin{figure}
\centering
\includegraphics[scale=0.165]{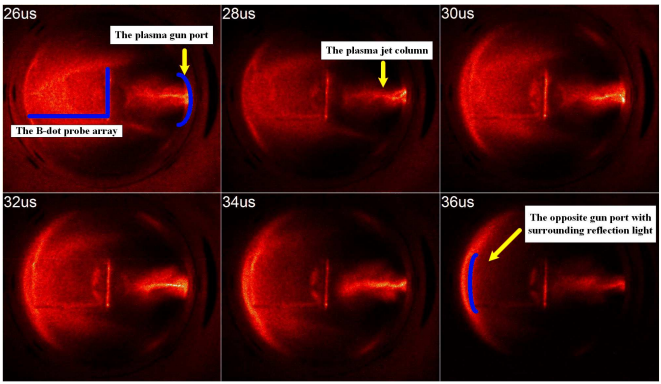}
\caption{(Color online). Plasma-jet images showing the central column becoming helical and a growing kink instability (shot no.\ 001050714, 2-$\mu$s interframe time). The 3D B-dot probe array (placed 25~cm away from the plasma gun port), the plasma gun port, and the jet body are labeled.}
\label{fig:vacjet}
\end{figure}

The gun is typically operated with a discharge current~$I_{gun}$ = 60--100~kA and a 6.0--10~kV charge voltage on the capacitor bank.~Experimental results to date have centered around time-resolved global imaging and probe measurements of the plasma-jet evolution. The images were taken using a Hadland Ultra UHSi 12/24 multiple-frame CCD camera, which takes up to twelve images per plasma discharge.~The camera view is through a 39-cm~$\times$~18.5-cm observation window at the end of the chamber (labeled in Fig.~\ref{fig:overview}) such that the gun port appears on the right-hand side of each image.~The exposure time of each frame is 500~ns and the interframe time is set typically to 2.0~$\mu$s, which is of the order of an Alfv\'en transit time. False color is applied to the images for ease of viewing.~Typically, the plasma is observed in unfiltered visible light.~Although the exact relationship between emission intensity and plasma density is complex, we use the light emission (allowing for a generous uncertainty) as a proxy for inferring the radius and length of the bulk mass of the jet. We allow for a conservative $\pm 50$\% error bar in the inferred radius and length to calculate the safety factor $q$.~The physics conclusions remain valid over the entire range of the uncertainty. 

A double Langmuir probe and a magnetic-probe array have been inserted into the chamber~(as indicated in Fig.~\ref{fig:overview})~to measure plasma parameters.~Typical plasma-jet densities, electron temperatures and approximate peak magnetic field are $10^{20}$~m$^{-3}$, 10~eV, and 0.1~T respectively.~The jet propagates from the gun to the opposite wall of the chamber at near the local Alfv\'en velocity (about 3.5$\times10^4$ m/s).~For such a dynamic system, the jet characteristics exhibit very robust, reproducible measurements for a wide parameter regime. Each setup has explored hundreds of shots which show similar jet dynamics.~The shot-to-shot standard deviation in the magnetic-field measurements over a large shot sample is about 15\%.

First, the coaxial plasma gun is operated to launch a collimated magnetized plasma jet into the vacuum chamber without a background magnetic field, as shown in Fig.~\ref{fig:vacjet}.~As the length of the jet increases, the~$n = 1$~kink instability develops.~The observed kinks in the plasma jet column have been analyzed and examined using Kruskal-Shafranov theory of current-driven instabilities, consistent with previous work.\cite{hsu2002laboratory,hsu2003experimental,hsu2005jets}~The safety factor and condition for instability is given by

\begin{equation} \label{eq:1}
q=\frac{2\pi aB_{z}}{lB_{\varphi }}<1,
\end{equation}
where $\mathit{a}$ and $\mathit{l}$ are the jet radius and length, respectively.

\begin{figure}
\centering
\includegraphics[scale=0.12]{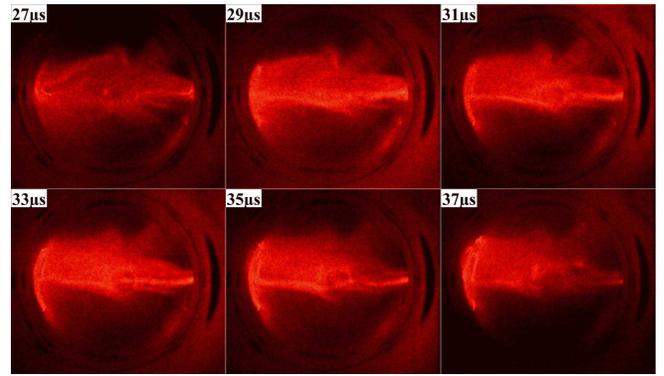}
\caption{(Color online). A more-stable plasma jet column is observed (shot no.\ 008050114, 2-$\mu$s interframe time) when injecting the plasma jet into a 500-G magnetic field transverse to the direction of jet propagation}
\label{fig:bgjet}
\end{figure}

Then, under the same operational settings of the coaxial gun, the plasma-jet is launched into a 500-G background magnetic field (transverse to the direction of jet propagation). The jet evolution for this case is shown in~Fig.~\ref{fig:bgjet}. The images indicate that~$\left(1\right)$~the plasma jet penetrates into the background magnetic field and~$\left(2\right)$ a more-stable jet is formed during the injection. As inferred from light emission in Fig.~\ref{fig:bgjet}, the jet length is on the order of 50 cm, which is much longer than the approximately 10-cm length observed for the vacuum case in~Fig.~\ref{fig:vacjet}. The jet life-time is~$\sim 3t_{Alfv\acute{\mathrm{e}}n}$~(where~$t_{Alfv\acute{\mathrm{e}}n} \equiv l/V_{Alfv\acute{\mathrm{e}}n}$), which is prolonged compared to the vacuum case for which the plasma-jet lifetime is $\sim t_{Alfv\acute{\mathrm{e}}n}$. Comparing the experimental settings for these two cases, the only difference is the presence of the background magnetic field in the main vacuum chamber.

Theoretical analysis has shown that for a plasma jet to penetrate into a transverse magnetic field, the kinetic-energy density of the plasma jet must overcome the magnetic energy density of the background magnetized plasma.\cite{brown1990current,zhang2016experimental}~This requirement may be stated as~$ \rho_{jet} {V_{jet}^{2}} /2 \geq  {B_{0}^{2}} /2 \mu_{0}$ , where $\rho_{jet}$ is the plasma jet mass density, $V_{jet}$ is the plasma jet velocity, and $B_{0}$ is the background magnetic field strength.~In this experiment, substituting jet density $n=10^{20}$~m$^{-3}$, $M_{argon} = 6.62\times10^{-26}$~kg, and $V_{jet} = 3.5\times10^{4}$~m/s into $\rho_{jet} {V_{jet}^{2}} /2$, one obtains $P_{jet}= 4.1\times 10^{3}$~Pa.~For ${B_{0}^{2}} /2 \mu_{0}$, using $B_{0}=500$~G, one obtains $P_{B} = 1.0\times 10^{3}$~Pa. Thus, $P_{jet}$ is greater than $P_{B}$,~which means that the plasma jet can propagate into the transverse field of the background plasma.

\begin{figure}
\centering
\includegraphics[scale=0.5]{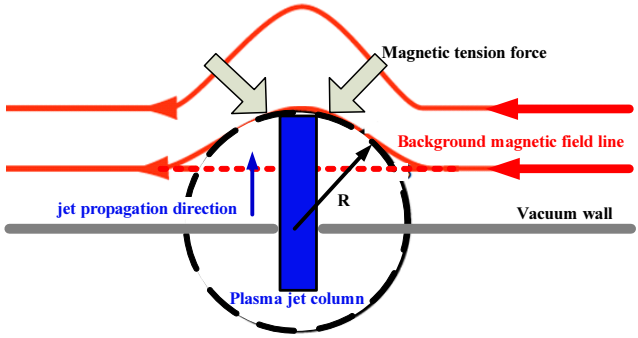}
\caption{(Color online).The plasma jet propagates into the transverse background magnetic field. The advected background magnetic field line is curved with the movement of the plasma jet, causing a tension force on the jet column.}
\label{fig:theory}
\end{figure}

The magnetic Reynolds number, which represents the ratio between magnetic advection and magnetic diffusion, is $R_{M}= \mu_0lV/\eta \sim~100$, where the characteristic length $l=0.5$~m, the characteristic velocity $V=3.5\times 10^4$~m/s, and the magnetic diffusivity $\eta=21.4$~m$^2$/s (the plasma-jet conductivity $\sigma=1.7\times 10^{7} $~S/m, using $T_{e}=10$~eV). The value suggests that magnetic diffusion is relatively unimportant on the length scale \textit{l}. The background magnetic field lines are then advected with the plasma jet propagation, as indicated in~Fig.~\ref{fig:theory}.~The curved magnetic field induces a magnetic tension force to act on the plasma jet as:

\begin{equation} \label{eq:2}
F_{tension}=\frac{B_{0}^{2}}{\mu _{0}}\cdot  \frac{\hat{R}}{R}~,
\end{equation}
where $\hat{\textit{R}}$ is the unit vector pointing from the magnetic field line to the center of curvature, and $\textit{R}$ is the radius of curvature of the field line labeled in Fig.~\ref{fig:theory}.

\begin{figure}
\centering
\includegraphics[scale=1.1]{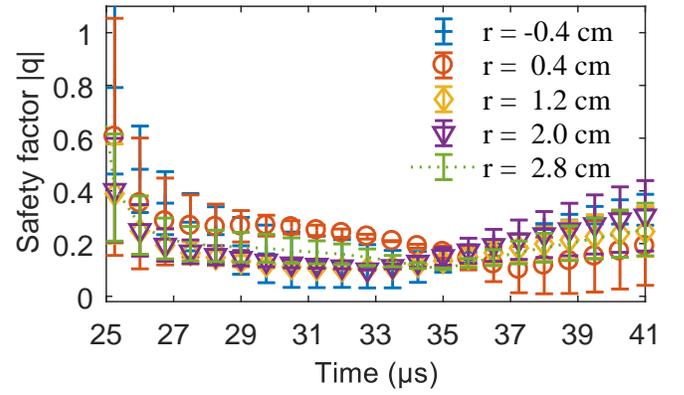}
\caption{(Color online) For the case with background magnetic field, safety factor $q$ vs.\ time at different radii with respect to the jet axis ($r=0$ cm), calculated based on the Kruskal-Shafranov criterion [Eq.~\ref{eq:1}] with $\pm$ 50\% error in inferring jet length and radius from light emission}
\label{fig:safety}
\end{figure}

\begin{figure}
\centering
\includegraphics[scale=0.79]{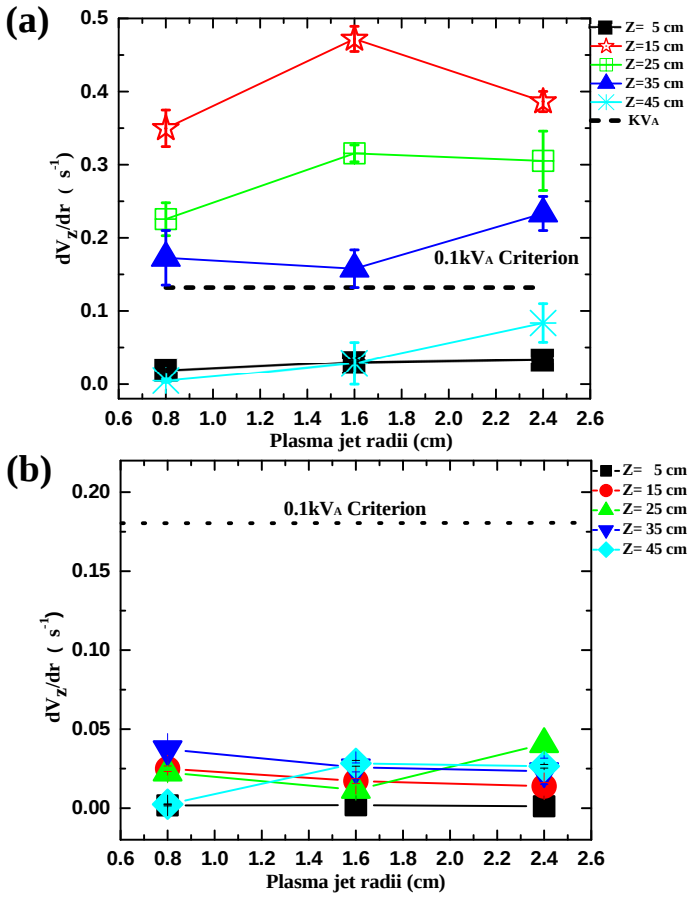}
\caption{(Color online) Top (a): the spatial velocity shear is calculated for plasma-jet injection into the transverse field of the background plasma. A positive velocity shear is obtained above the threshold value (indicated by the dash-dot line). Bottom (b): the calculated spatial-velocity shear for the plasma jet launched into vacuum, well below the sheared-flow stabilization threshold.}
\label{fig:shear}
\end{figure}

It is clear from Fig.~\ref{fig:theory}~that the background magnetic tension force is against the axial~$\mathit{J\times B}$~force which is from the main capacitor-bank discharge current, and this configuration induces a sheared axial flow of the plasma jet. The Kruskal-Shafranov criterion has been employed for the jet-instability analysis in this case.~Figure~\ref{fig:safety}~shows the calculated safety factor, $\textit{q}$, from~Eq.~\ref{eq:1}, using the measured magnetic-field data and plasma-jet dimensions from the CCD camera images. As shown in Fig.~\ref{fig:safety}, with $\pm$ 50\% error in inferring jet length and radius from light emission, the calculated~\mbox{\lq localized\rq}~safety factors, where the B-dot probe is placed, are below unity. Based on the Kruskal-Shafranov criterion, the plasma jet should experience the kink instability, which will deform and break the jet column. However, the image data shown in Fig.~\ref{fig:bgjet} indicates a globally more-stable plasma jet column. Thus, the Kruskal-Shafranov criterion is not sufficient in this case to assess the plasma kink stability. The axial sheared flow, which is caused by the background magnetic-tension force, contributes to the global jet-stabilization process.
The effect of sheared flow on current-driven MHD instabilities has previously been investigated theoretically and demonstrated experimentally.\cite{shu1995sheared,shumlak2001evidence,shumlak2003sheared,golingo2005formation,shumlak2006plasma,shumlak2012sheared,shumlak2017increasing}~Linear MHD calculations show that a sheared axial flow has a stabilizing effect on the kink mode, while a uniform axial flow has no effect on the instability growth. The main prior conclusion is that an axial plasma flow with a linear shear of~$dV_{z}/dr>0.1kV_{A}$~is required for jet stabilization, where~$\textit{k}$~is the axial wave number and~$\textit{V}_\textit{A}$~is the Alfv\'en velocity.\cite{shumlak2003sheared}
The radial profile of the plasma axial flow velocity in this experiment is calculated based on the time delay between the initial rise of the magnetic field measured by the magnetic probe array, which is placed at various positions along the jet propagation axis in the main chamber at 10-cm intervals.~After the radial velocity profile obtained at different positions in the vacuum chamber, the axial sheared flow $dV_{z}/dr$  is determined and the result is plotted in Fig.~\ref{fig:shear} (top).~As shown in this plot, for the main region of the chamber ($z=15$--35~cm), the magnitude of the axial shear is above the threshold value $0.1kV_{A}$, consistent with the theoretical analysis of the axial sheared flow stabilization. As a comparison, the axial flow shear for plasma-jet injection into vacuum is also calculated and the result is shown in Fig.~\ref{fig:shear} (bottom). There is no flow shear for plasma-jet propagation into vacuum, and the uniform axial flow has no stabilizing effect on the kink instability.

In summary, we have demonstrated the phenomenon of emergent kink stability of a magnetized plasma jet by injecting the jet into a transverse background magnetic field. After the plasma jet leaves the plasma-gun muzzle, a background magnetic-tension force arises because the jet bends the transverse background magnetic field, and a global more-stable jet column is observed from the image data. Evaluation of the Kruskal-Shafranov criterion based on the measured safety factor indicates that the jet should be kink unstable. However, the evolution of the axial velocity profile shows an axial velocity shear that is above the theoretical threshold kink-stabilization threshold $0.1kV_{A}$, resulting in the kink stabilization of the plasma jet.

The authors thank Dr.~Glen Wurden for loaning the CCD camera and Dr. Kevin Yates for assisting with its setup. This material is based upon work supported by the National Science Foundation under Grant no. AST-0613577 and the Army Research Office under award no. W911NF1510480.

\end{document}